\documentclass[letterpaper]{article}

\usepackage{natbib,alifeconf}  
\usepackage{url}
\usepackage{mathabx}

\usepackage{hyperref}
\title{Step Size is a Consequential Parameter in Continuous Cellular Automata}

\author{Q. Tyrell Davis and Josh Bongard \thanks{$^\ast$ University of Vermont, [qdavis, jbongard]@uvm.edu} }

\begin{document}

\maketitle


\section{Introduction}

In the 1960s, John H. Conway developed a zero-player game with simple rules. This Game of Life, a cellular automaton (CA), has had a seminal impact on the study of complex systems, computation, and art.
Conway's Life followed John von Neumann's 29-state CA \citep{neumann1966}, and Life's impact on popular as well as academic imagination is unique, seeded by a 1970 article in Scientific American by Martin Gardner \citep{gardner1970}.

Subsequent decades saw increasing diversity of CA research and applications.
The Life framework: totalistic CA based on a Moore neighborhood and cell states, supports 262,144 different rulesets\footnote{Life-like CA are defined by \textbf{B}irth and \textbf{S}urvival rules; each may contain any or all of the possible Moore neighborhood sum values 0 through 8, yielding $2^9 \cdot 2^9 = 2^{18} = 262,144$ possible rules. At each step, cells with a neighborhood sum in their S rules remain unchanged, become 1 with a sum in B; all other cells become 0.}
CA have since been developed with larger neighborhoods \citep{evans2001, pivato2007}, higher dimensions \citep{bays1987, chan2020}, and evolving rules \citep{mccaskill2019}, to name but a few examples among many. In this work we are concerned with continuous CA.

Rudy Rucker developed a continuous CA framework in the 1990s called CAPOW \citep{rucker2003}. Later, Stephen Rafler developed a continuous CA, SmoothLife, and discovered a persistent glider therein \citep{rafler2011}. Recently, Bert Chan described the discovery of many persistent patterns in his Lenia framework \citep{chan2019, chan2020}, and encoding CA updates in continuously valued neural networks has been applied to models for growth \citep{mordvintsev2020}, image recognition \citep{randazzo2020}, and control \citep{variengien2007}. Continuous CA can be described as:

\begin{equation}
    A(t+dt) = A(t) + dt \cdot f(A(t))
    \label{eqn:general_cca}
\end{equation}

where $A$ represents cell states, $t$ is unitless time, $dt$ is step size, and $f(\cdot)$ represents some function dependent on cell states\footnote{$f(\cdot)$ typically includes a neighborhood operation ({\itshape e.g.} convolution) and an arbitrary update function.}

\begin{figure}[t]                                                      
\begin{center}                                                          
  \includegraphics{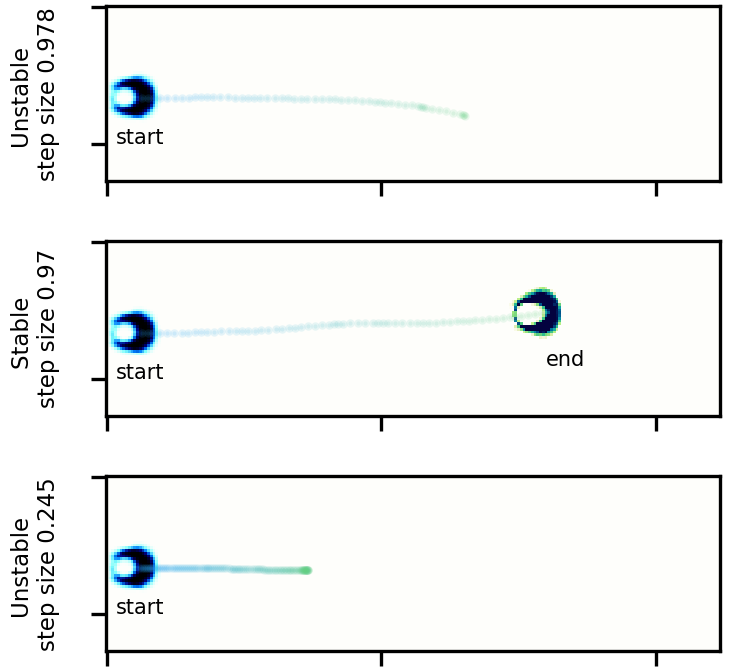}
    \caption{Stability of a minimal glider in Lenia's {\itshape Scutium gravidus} ruleset spans from approximately $dt = 0.25$ to $0.97$. Kernel parameters $(\mu_K, \sigma_K) = (0.5, 0.15)$ , update parameters $(\mu_G, \sigma_G) = (0.283, 0.0369)$.}
  \label{fig:scutium_trajectory}
\end{center}        
\end{figure}  

Typically receiving little scrutiny, $dt$ can have important effects. Persistent patterns require step size to be neither too large nor too small, and multiple patterns may exist their own $dt$ ranges within an otherwise identical CA rule set (Figures 2 and 3).
Step size can also lead to qualitatively different behavior in CA. Varying step size from 0.0125 to 0.13 in Figure \ref{fig:frog_trajectory} yields diverse movement types including hopping, meandering, and corkscrewing. While not a perfect analogy, the behavioral repertoire in Figure \ref{fig:frog_trajectory} seems to have more in common with a robot in a conventional physics simulation changing from jumping to pirouette movement patterns than with the expected catastrophic failure (or tedious slow-down) caused by an inappropriate time-step in a conventional differential equation-based numerical physics simulation. Supporting resources for this project are open-source \footnote{Links to animations, notebooks, and code used in this project are consolidated at \url{https://rivesunder.github.io/yuca}}



\section{Pattern stability and step size}

An intuitive consequence of poor step size choice is that patterns become unstable when the step size is too large, but this also occurs when step size is too small. This defies the expectation that in simulations of physical system we typically expect greater accuracy as the step size approaches zero. What would be considered systematic error in simulating billiard ball trajectories is essential for self-organization by some CA patterns. 

A minimal glider {\itshape Scutium gravidus} in the Lenia framework, cousin to the SmoothLife glider \citep{rafler2011}, is stable between approximately $dt = 0.245$ to $0.978$, smaller or larger step sizes lead to the glider vanishing. Sample trajectories at both extremes are shown in Figure \ref{fig:scutium_trajectory}.

\begin{figure}[t]                                                      
\begin{center}                                                          
  \includegraphics{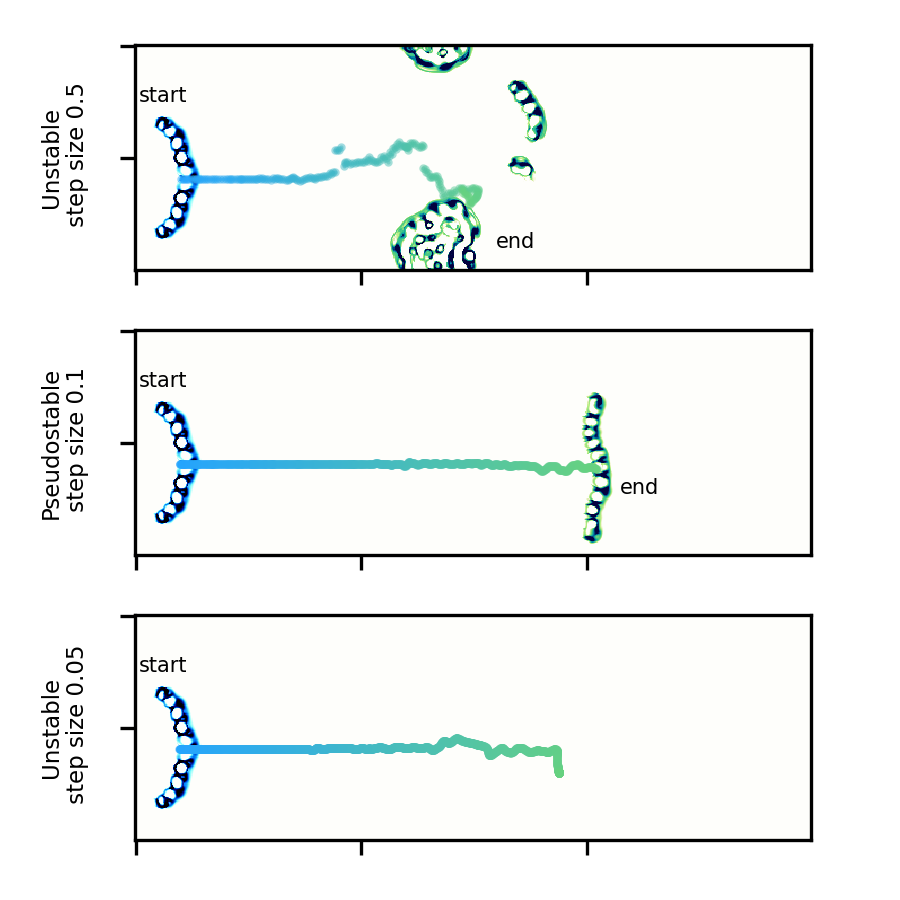}
    \caption{This wide glider pattern is pseudostable around $dt=0.1$, but tends to change size and shape (sometimes with explosive growth) at higher $dt$, and to disappear at lower $dt$.  Kernel parameters $(\mu_K, \sigma_K) = (0.5, 0.15)$ , update parameters $(\mu_G, \sigma_G) = (0.283, 0.0369)$.}
    \label{fig:superwide_scutium}
\end{center}        
\end{figure}

Another pattern operating under the same neighborhood and update rules, a wide glider, is most stable at much lower step sizes around 0.1. Larger or smaller step sizes yield unconstrained growth or a vanishing pattern, respectively (Figure \ref{fig:superwide_scutium}).  A step size of 0.1 is pseudo-stable for this pattern, but is sensitive to initial conditions (including grid dimensions, pattern placement, and floating point precision) and can eventually ($\sim$1000s of steps) become unstable. 

\section{Pattern behavior and step size}

\begin{figure}[!ht!]                                                              
\begin{center}                                                              
  \includegraphics{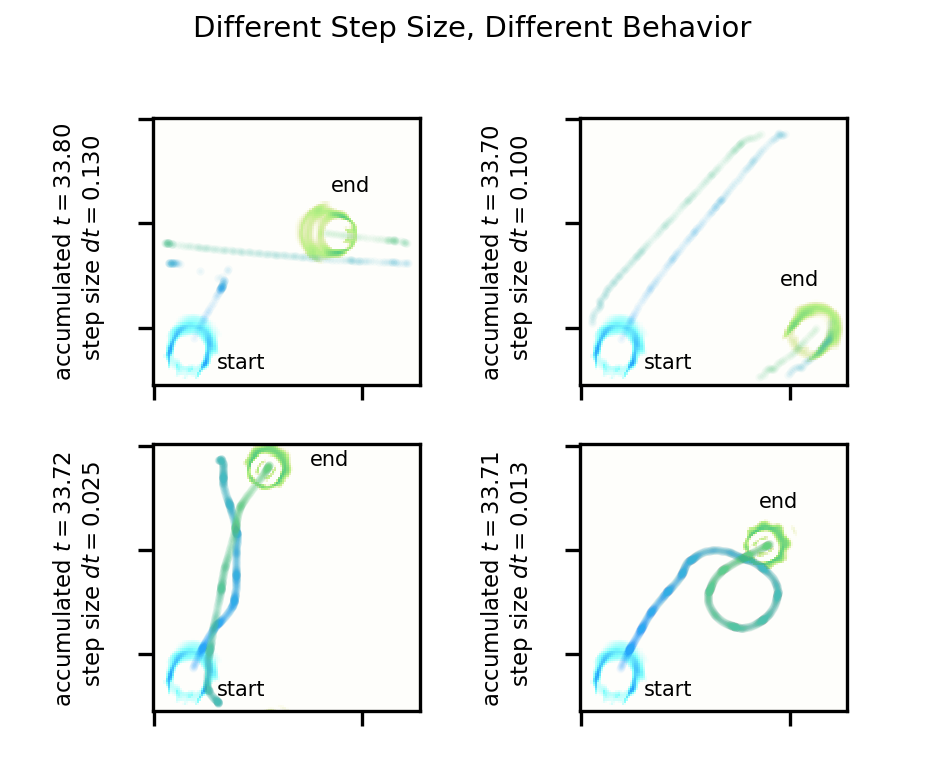}
  \caption{Identical starting conditions, here a mobile ``frog" pattern, follow different trajectories with qualitatively different behavior under CA rules differing only in step size $dt$. Note the CA grid represents the surface of a torus: trajectories reaching one edge of the grid reappear at the opposite edge.}
  \label{fig:frog_trajectory}
\end{center}        
\end{figure} 

In addition to pattern instability, step size can lead to qualitatively different behavior. Figure \ref{fig:frog_trajectory} demonstrates behavioral diversity solely by changing step size. At the ``natural" step size of 0.1, it moves in a ``hopping" motion. Relatively large step sizes ($\approx 0.13$) bring the pattern to the edge of stability, with occasional explosive growth (responsible for a sharp turn in trajectory), and also occasional larger, surging hops. At step size 0.025 the pattern travels in a meandering trajectory and a step size of  0.0125 leads to corkscrew trajectories and a spiky morphology. The CA in Figure \ref{fig:frog_trajectory} uses the Glaberish CA framework \citep{davis2022}, splitting the update function into persistence and genesis functions dependent on cell state, with a neighborhood kernel of mixed Gaussians with parameters $(\mu_K, \sigma_K) = [(0.0938, 0.033), (0.2814, 0.033), (0.469, 0.033)]$ with weights $[0.5, 1.0, 0.667]$, a Gaussian genesis function $(\mu_g, \sigma_g) = (0.0621, 0.0088)$, and persistence function $(\mu_p, \sigma_p) = (0.2151, 0.0369)$.

\section{Conclusions}

This work demonstrates that step size is a consequential parameter in continuous CA, affecting pattern stability and qualitative behavior. This is in marked contrast to remarks in \citep{chan2019}, which, noting the resemblance of the Lenia update to Euler's method, suggested that decreasing step size asymptotically approaches the ideal simulation of a Lenia pattern, {\itshape Orbium}. This work demonstrates that for several patterns a lower step size does not entail a more accurate simulation, but different behavior or potential patterns entirely. Therefore, step size should be considered for optimization and learning with CA, \textit{e.g.} to develop patterns with agency to avoid obstacles \citep{hamon2022} or for training neural CA such as in\citep{mordvintsev2020, randazzo2020, variengien2007}.



\subsection{Funding} This work was supported by the National Science Foundation under the Emerging Frontiers in Research and Innovation (EFRI) program (EFMA-1830870).

\newpage
\footnotesize
\bibliographystyle{apalike}
\bibliography{example} 

\end{document}